\documentclass[%
 aip,
 amsmath,amssymb,
reprint,  
]{revtex4-1}
\usepackage{graphicx}
\usepackage{dcolumn}
\usepackage{bm}
\usepackage{chemmacros}
\usepackage[utf8]{inputenc}
\usepackage[T1]{fontenc}
\usepackage{mathptmx}
\usepackage{etoolbox}
\usepackage{amsmath}
\usepackage{comment}
\newcommand{\angstrom}{\textup{\AA}}

\newcommand{\DefineAuthor}[2]{%
  \expandafter\newcommand\csname #1note\endcsname[1]{%
    \textbf{\textcolor{#2}{\textbf{#1:} ##1}}}%
  \expandafter\newcommand\csname #1\endcsname[1]{
    \textbf{\textcolor{#2}{##1}}}
  \expandafter\newcommand\csname #1cancel\endcsname[1]{%
    \textbf{\textcolor{#2}{\sout{##1}}}}%
  \expandafter\newcommand\csname #1change\endcsname[2]{%
    \textbf{\textcolor{#2}{\sout{##1} ##2}}}%
  \newenvironment{#1text}{\color{#2}}{\color{black}}
}
\definecolor{dartmouthgreen}{rgb}{0.05, 0.5, 0.06}
\makeatletter
\def\@email#1#2{%
 \endgroup
 \patchcmd{\titleblock@produce}
  {\frontmatter@RRAPformat}
  {\frontmatter@RRAPformat{\produce@RRAP{*#1\href{mailto:#2}{#2}}}\frontmatter@RRAPformat}
  {}{}
}%
\makeatother
\begin{document}

\preprint{AIP/123-QED}

\title{Are nonequilibrium effects relevant for chiral molecule discrimination?}

\author{Federico Ravera}
 \affiliation{Department of Electronics and Telecommunications, Politecnico di Torino, 10129 Torino, Italy}
\author{Leonardo Medrano Sandonas}%
\affiliation{Institute for Materials Science and Max Bergmann Center of Biomaterials, TUD Dresden University of Technology, 01062 Dresden, Germany}%
\author{Rafael Gutierrez}
\affiliation{Institute for Materials Science and Max Bergmann Center of Biomaterials, TUD Dresden University of Technology, 01062 Dresden, Germany}%
\author{Mariagrazia Graziano}
\affiliation{Department of Applied Science and Technology, Politecnico di Torino, 10129 Torino, Italy}%
\author{Gianaurelio Cuniberti}
\affiliation{Institute for Materials Science and Max Bergmann Center of Biomaterials, TUD Dresden University of Technology, 01062 Dresden, Germany}%
 \email{leonardo.medrano@tu-dresden.de, rafael.gutierrez@tu-dresden.de, gianaurelio.cuniberti@tu-dresden.de.}

\date{\today}

\begin{abstract}
Sensing and discriminating between enantiomers of chiral molecules remains a significant challenge in the design of sensor platforms. In the case of chemoresistive sensors--where detection relies on changes in electrical response upon analyte adsorption--the sensor substrate is typically functionalized with chirality-sensitive molecular receptors. 
In this computational study, we investigate whether a chirality-blind substrate, such as a graphene nanoribbon, is still capable of discriminating between enantiomers. 
To this end, we employ a density-functional parametrized tight-binding method combined with nonequilibrium Green’s functions. For a small set of chiral amino acids, we demonstrate that accounting for the nonequilibrium response of the device leads to significant differences in the electrical currents of enantiomeric pairs of the order tens of nanoamperes. This effect is further amplified when structural fluctuations of the device's active region are considered ($\approx$1-2 $\mu$A).
Moreover, we propose new quantum-mechanical quantities for enantioselective discrimination in molecular sensors, with an emphasis on binding features and property-property correlations. 
Therefore, our work demonstrates the significance of nonequilibrium effects in chiral discrimination, laying the foundation for future investigations addressing the design of chiral molecular sensors.
\end{abstract}

\maketitle

\section{\label{intro} Introduction}

Chirality is a fundamental characteristic of life and a very common phenomenon in nature. A chiral object is one that cannot be superimposed on its mirror image, with the two mirror images being referred to as enantiomers.\cite{Kumar2009}
Chiral molecules are ubiquitous in pharmaceutical applications,\cite{D4RA05694A} food additives,\cite{ALVAREZRIVERA2020115761} and in biological systems,\cite{sym14030460,Ma2023} among others. 
Interestingly, the enantiomers of a molecular compound can exhibit distinct behaviors, such as differences in transport mechanisms, metabolic pathways, and biochemical activities.\cite{Ribeiro2021,Pu2019} 
For example, one enantiomer in a racemic drug mixture may be toxic, potentially causing undesired side effects or even death at high doses.\cite{Tenconi1994} This is the case, e.g., of the enantiomers of chiral thalidomide, which have dramatically different impacts on human beings: whereas R-thalidomide helps to alleviate morning sickness in pregnant women, S-thalidomide may induce severe birth defects.
Accordingly, it is critical to develop effective analytical tools for enantioselective discrimination.

Over the years, a variety of experimental techniques have been developed for this purpose, including circular dichroism, electrochemical impedance spectroscopy, fluorescence spectroscopy, surface-enhanced Raman scattering,  gas chromatography-mass spectrometry, and nuclear magnetic resonance. \cite{Qin2020,Wang2016,Qin2020} 
However, many of these methods involve the use of specialized equipment, such as circularly polarized light, or require the preparation of chiral selectors.  
Of particular interest for the purpose of the current investigation are chemoresistive gas sensors for chiral recognition \cite{Hulanicki1991,Rath2023,Luo2023,doi:10.1021/acs.jpclett.1c03106,ZOR2017410,chemosensors10080308,doi:10.1021/acs.analchem.3c00669,doi:10.1021/acsanm.0c00389,CHEN2023117562} that operate on the basis of changes in the electrical resistance when exposed to a target gas. 
For providing an efficient chiral discrimination, these sensing devices usually require the design of specifically tailored chiral receptors, which provide a chiral interface capable of non-covalently binding the (chiral) analytes as well as distinguishing between enantiomeric conformations.  
The microscopic picture underlying this approach involves a complex interplay of interrelated effects. Chiral molecules interact with chirality-sensitive receptors functionalizing the sensor substrate --primarily through van der Waals interactions, hydrogen bonding, and electrostatic forces--leading to charge redistribution that alters the electronic structure of both the receptor and the substrate. These modifications, in turn, can influence the electrical response of the substrate, enabling chiral discrimination.

From a computational modeling perspective, studying gas sensing --especially chiral discrimination--in chemoresistive sensor setups is highly challenging. It demands an accurate treatment of noncovalent interactions, charge transfer effects, a reliable configurational sampling, and charge transport.
Unlike well-established computational electron transfer and transport approaches in molecular junctions,\cite{10.1093/oso/9780198529798.001.0001,Sadeghi_2018,Hirsbrunner_2019,PhysRevB.87.245407,PhysRevB.87.085422,PhysRevB.87.245407,doi:10.1021/acsnanoscienceau.4c00041,Solomon2010,doi:10.1021/acs.jpcc.0c08957} where the molecular system of interest is directly involved in electrical conduction, in  chemoresistive sensing devices (chiral) gas molecules influence the electronic transport on the substrate only indirectly through the previously mentioned charge density reorganization. Their effect can thus be viewed as that of quasi-localized nanoscale electrostatic gates, making their impact on the electrical response more subtle. 
We also remark that there are computational studies probing quantum transport in molecular systems with structural chirality; \cite{https://doi.org/10.1002/anie.202218640,doi:10.1021/acs.jpcc.0c09364,García-Inglés_2025,doi:10.1021/acs.jpclett.3c02702} however, the employed transport two-terminal setups are conceptually different from the problem of chemoresistive chiral sensing,   probing  charge transport through the molecular systems themselves. Besides these studies, we are not aware of any computational studies of chiral discrimination in chemoresistive sensing.

\begin{figure}[t!]
 \centering
 \includegraphics[width=\columnwidth]{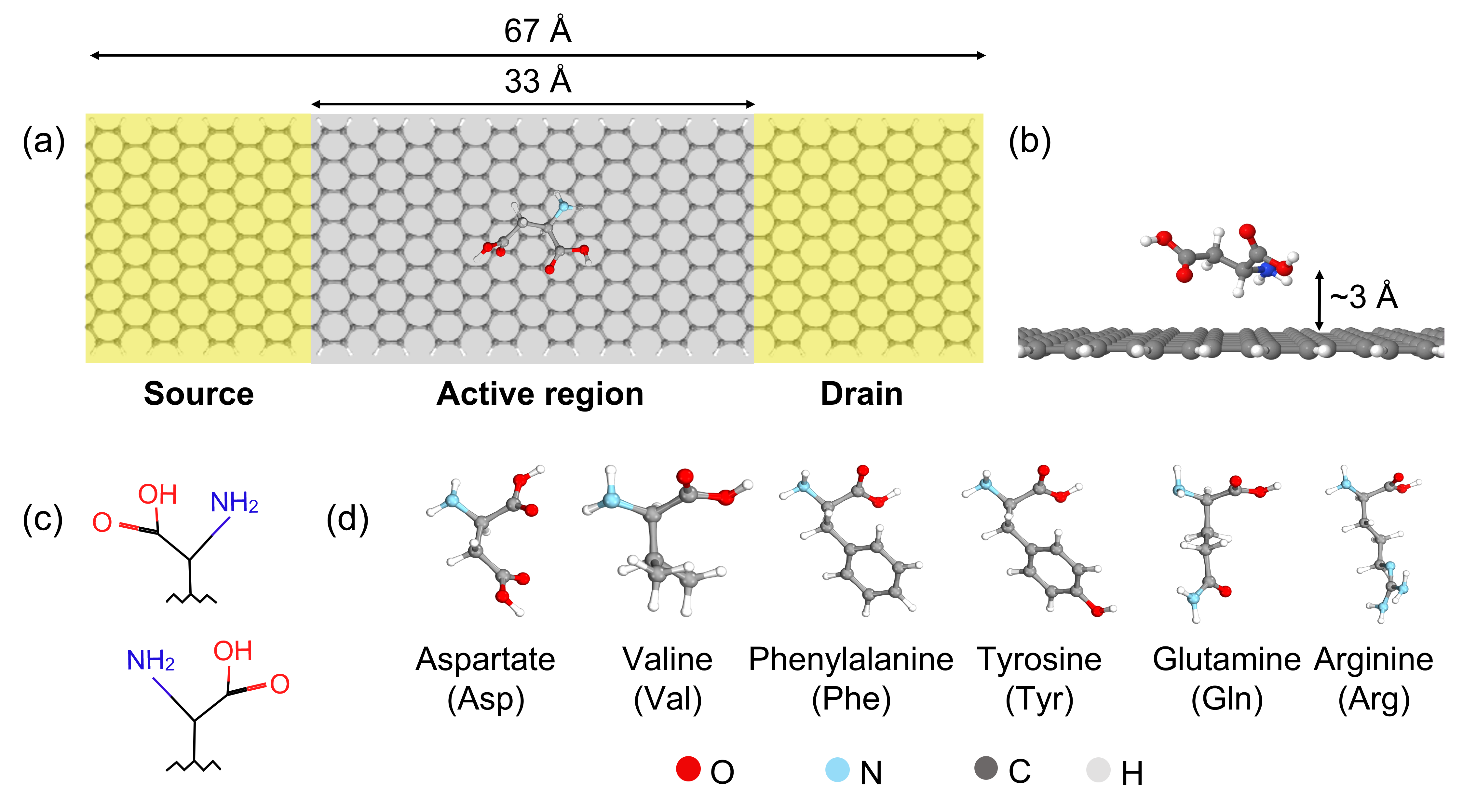}
 \caption{(a) Top view of the two-terminal setup used for the charge transport calculations. A finite size domain of an infinite armchair graphene nanoribbon (AGNR15) is considered as the sensing substrate ("active region") with the two semi-infinite domains act as electronic reservoirs. A given chiral molecule interacts with the "active region". The domains highlighted as "source" and "drain" are the regions where the potential difference is applied; they are not structurally relaxed or affected by the presence of the molecules. (b) Side view of the "active region". The molecules are initially placed at a typical van-der-Waals distance of $\sim 3 \angstrom$ above the AGNR substrate. (c) Schematic representation of the enantiomers, highlighting the common chiral center for all amino acids. (d) The set of amino acids used in this investigation (only the D-enantiomers are shown for simplicity).}
 \label{fig:1}
\end{figure}

In the present work, we investigate the following question: Can enantiomers be identified without the need for tailored molecular receptors, simply by allowing chiral molecules to interact directly with a chirality-“blind” substrate? 
To explore this, we select a free-standing graphene nanoribbon as the sensing material \cite{JOHNSON2021113245} and expose it to a set of chiral amino acids recently used in a theoretical study   on enantiospecificity in NMR experiments\cite{Georgiou2024}, see scheme in Fig. \ref{fig:1}.
Computationally, we combine a Density-Functional Tight-Binding\cite{doi:10.1021/ct100684s} based electronic structure and quantum Molecular Dynamics with Nonequilibrium Green functions.\cite{Ryndyk2009}  The main observable of interest is the electrical current flowing through the nanoribbon and its variations when  L- and D-enantiomers interact with the substrate. 
Our results emphasize two key findings: (i) the linear (equilibrium) conductance exhibits no enantioselective features; however, a  non-equilibrium transport calculation at finite voltages reveals significant differences in electrical current between enantiomers for a given spatial orientation relative to the graphene substrate; (ii) these current differences become even more pronounced when the conformational dynamics of the chiral molecules is incorporated into the transport calculations within an adiabatic approximation.
Our results suggest that non-equilibrium transport effects and the need for a meaningful sampling of the conformational space are crucial factors for understanding chiral sensing in chemoresistive devices.


\begin{figure*}[ht!]
 \centering
 \includegraphics[width=\textwidth]{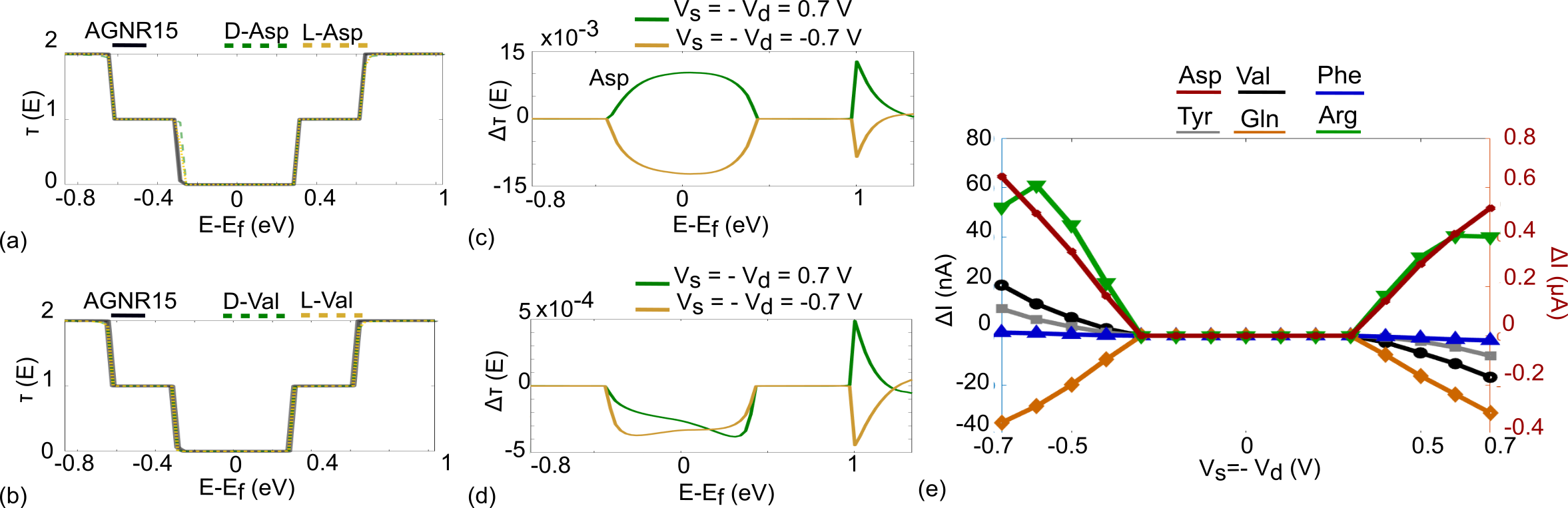}
 \caption{Zero-bias transmission spectrum, $\tau(E, V_{DS}=0)$,  of a bare graphene nanoribbon AGNR15 (black) and the corresponding spectrum of the graphene nanoribbon interacting with structurally optimized (a)  D-Asp (green) and L-Asp (yellow), and (b)  D-Val (green) and L-Val (yellow).
Transmission spectrum difference $\Delta \tau = \tau_L -\tau_D$ for (a) Asp at $V_S = -V_D = 0.7$ V (green) and $V_S = -V_D = -0.7$ V (yellow) and (b)  Val at $V_S = -V_D = 0.7$ V (green) and $V_S = -V_D = -0.7$ V (yellow).
The energy values were shifted relative to the Fermi energy $E_f$ of the transport setup.
(e) Variation of the difference in electrical current, $\Delta I = I_L - I_D$, as a function of applied voltage ($\Delta I-V$ curves) for the static conformations of all analyzed amino acids: $V_S = -V_D \in [-0.7, 0.7]$ V.}
 \label{fig:2}
\end{figure*}

\section{Computational methods}

\subsection{Chiral descrimination setup}

Our work examines the capabilities of a graphene nanoribbon as a chiral molecular sensor in equilibrium and nonequilibrium thermodynamic states, see Fig. \ref{fig:1}. 
To this end, we used a NEGF+DFTB approach, combining nonequilibrium Green functions (NEGF)\cite{Ryndyk2009}  with the  third-order semi-empirical density functional tight-binding (DFTB3) approach.\cite{doi:10.1021/ct100684s}
This methodology has been successfully used to investigate electron transport properties of a wide variety of low-dimensional systems such as molecular junctions,\cite{doi:10.1021/jacs.4c13531,doi:10.1021/jp512524z} two-dimensional materials, \cite{doi:10.1021/acs.jpcc.6b04969,PhysRevMaterials.6.114003} and nanotubes.\cite{Monavari2024}. 
Here, we consider an infinite armchair graphene nanoribbon with 15 armchair chains running along its length    (AGNR15), partitioned into three regions, corresponding to source (S), drain (D), and a central region; to the latter, the chiral molecules were allowed to interact with, thus defining the active region of the sensing device, see Fig.\ref{fig:1} for reference. 
To elucidate the potential of this setup for  chiral discrimination, we  studied the L- and D-enantiomers of six chiral amino acids with increasing size and complexity: aspartate, valine, phenylalanine, tyrosine, glutamine, and arginine. The initial conformations of these molecules were extracted from  Ref. \onlinecite{Georgiou2024}. While in general terms the definition of chirality for an arbitrary structure is not a simple issue,\cite{10.1063/5.0200716,PhysRevLett.133.268001} it does not pose any difficulties in our case, where only chiral centers are present in these small molecules.

Before geometry optimization of the AGNR-molecule systems, the chiral molecules were rigidly positioned at approximately 3 \(\text{\AA}\) over the active device region, with the longer dimension always parallel to the transport direction (see Fig. \ref{fig:1}(a)). 
The AGNR-molecule system was then  optimized using DFTB3\cite{hourahine2020dftb+} supplemented with a many-body treatment of van-der-Waals/dispersion interactions (MBD).\cite{mbd2012,mbd2014}
Our calculations require the inclusion of the MBD method due to its relevance in describing non-covalent interactions in complex systems that are not properly captured by the DFTB method. 
To compute the transport properties with the NEGF+DFTB approach, the Poisson equation was solved within a box with dimension $[40.0 \times 40.0 \times 30.0] \ \text{\AA}$ and a minimum grid spacing of 0.2 \text{\AA}. \cite{gaus2013parametrization}
Charge transport was evaluated with an energy state resolution set by a $\delta$ value of $10^{-4}$ eV. The Broyden mixing scheme, with a parameter of 0.05, was used for convergence. 
The key quantity to obtain the charge current is the energy and, for the nonequilibrium case, voltage dependent Landauer transmission $\tau(E,V_{DS})$, which is defined as:
\begin{eqnarray}
  \tau(E,V_{DS}) = \mbox{Trace}\left( G^{r}\Gamma_{L}G^{a}\Gamma_{R} \right) \ ,
\label{eq:Landauer1}
\end{eqnarray}
where $G^{r}(E,V_{DS})$ is a retarded Green's function of the active region computed as $G^{r}(E,V_{DS})=(EI - H - \Sigma^{r}_{L}(E,V_{DS})-\Sigma^{r}_{R}(E,V_{DS}))^{-1}$, the corresponding Hamiltonian matrix for that region is represented by $H$, $E$ is the energy of the electrons and $I$ is a  unit matrix with the dimension of the electronic space defined by $H$.
The broadening functions $\Gamma_{L/R}(E,V_{DS})=i\left[\Sigma^{r}_{L/R}(E,V_{DS})-\Sigma^{a}_{L/R}(E,V_{DS})\right]$ define the electronic bath spectral densities (semi-infinite source and drain AGNR), while the retarded/advanced self-energies $\Sigma^{r/a}_{L/R}(E,V_{DS})$ encode the electronic structure of both the semi-infinite electronic reservoirs and the reservoir-active region interface.
The transmission was calculated within the energy range -6.0 eV to -3.0 eV, with an energy step of 0.03 eV.
Based on the  voltage-dependent transmission function, the current-voltage ($I-V$) curve for each AGNR-molecule system was calculated by using the equation: \cite{Datta_2005_AtomToTrans}
\begin{equation} 
I(V_{DS})= \frac{2e}{h} \int_{-\infty}^{+\infty} \tau (E, V_{DS}) \left [ f_S(E) - f_D(E) \right] dE
\label{eq:Landauer}
\end{equation}
where $e$ is the elementary charge, $h$ Planck's constant, and $E$ is the electron energy. $f_S$ and $f_D$ are the Fermi-Dirac distributions for the source (S) and drain (D) electrodes, respectively. The current was evaluated for $V_{D} = -V_S \in [-0.7, 0.7]$ V with steps of 0.1 V. By using Eq.\ref{eq:Landauer1} we are assuming that only coherent transport is taking place, i.e., no coupling of the charges to  vibrational degrees of freedom is considered. 

\begin{figure*}[t!]
 \centering
 \includegraphics[width=\textwidth]{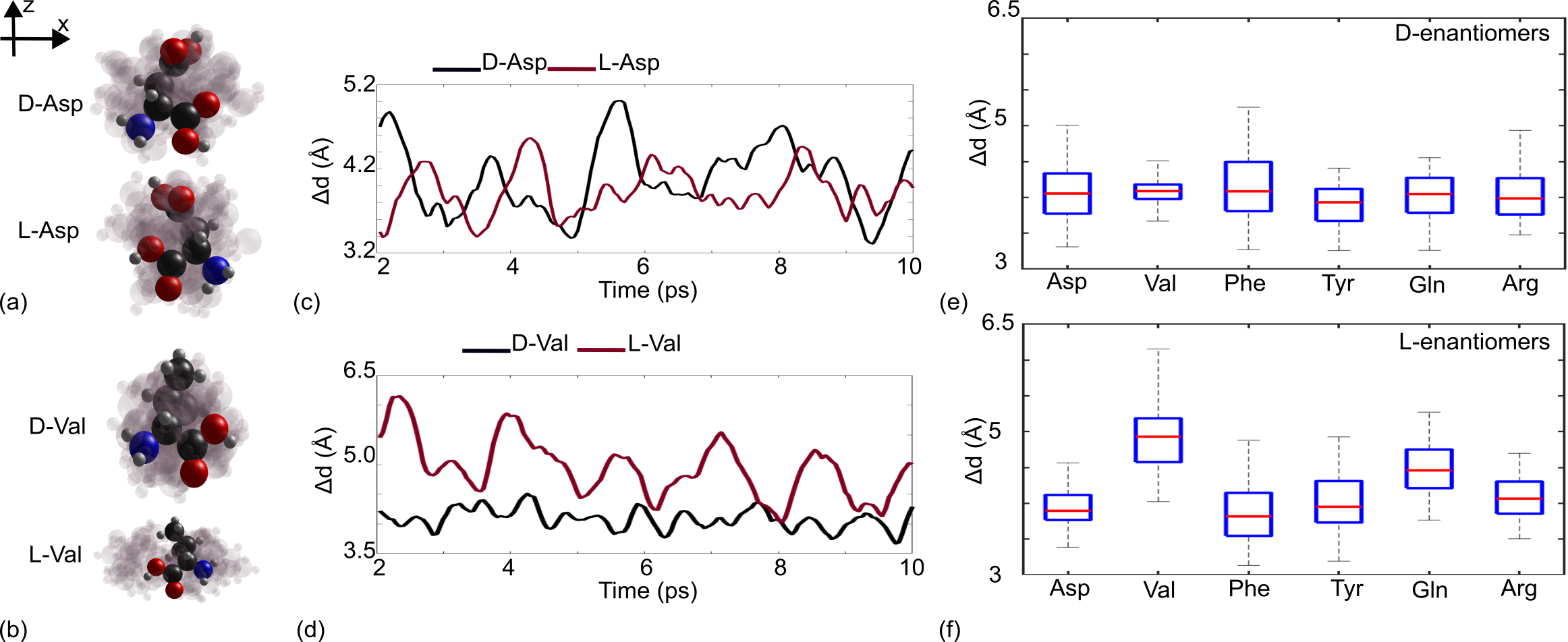}
 \caption{Illustration of the optimized structure (solid spheres) along with the nonequilibrium conformations (shadowed spheres) generated during the molecular dynamics (MD) trajectory for both enantiomers of (a) Asp and (b) Val.
Time dependence of the distance between the center of mass of the chiral molecule and AGNR15, $\Delta d(t)$, for both enantiomers of (c) Asp and (d) Val.
Boxplots of the $\Delta d$ values obtained during the MD trajectory for (e) D-enantiomer and (f) L-enantiomer of the six amino acids considered in this work.
%
}
 \label{fig:3}
\end{figure*}

\subsection{Structural dynamics and transport properties}\label{Sec:md}

To analyze the potential influence of structural fluctuations on the electron transport properties, molecular dynamics (MD) simulations at constant temperature were performed for each AGNR-molecule system (and for both L- and D-enantiomers) using the semi-empirical DFTB3 method with MBD corrections. Only the active device region was allowed to evolve in time, while the contacts were kept fixed. 
The MD simulations were run for 10 ps using a Nose-Hoover thermostat to maintain the temperature at 300 K within the NVT ensemble, with a time step of 0.5 fs.
From each MD trajectory, we extracted 160 nonequilibrium conformations from the last 8 ps in steps of 50 fs.
Then, the electrical currents for this set of  nonequilibrium conformations were calculated  using the NEGF+DFTB approach, and from them the corresponding ensemble averages were obtained. 
We remark that this approach can be seen as an adiabatic approximation, meaning that the time variable is a parameter labeling different configurations, and thus the conformational dynamics is not directly coupled to the charge transport process in a self-consistent way.  

Additionally, several parameters characterizing the interaction  between the AGNR and the chiral molecules along the MD trajectory were evaluated to better rationalize the changes in transport properties. 
In doing so, we have computed the binding energy of the system, $E_{\rm b}$, using the supramolecular approach: 
\begin{equation} \label{eq:Eb}
E_{\rm b}(t) = E_{\rm sys}(t) - E_{\rm AGNR}(t) - E_{\rm mol}(t),
\end{equation}
\noindent where $E_{\rm sys}(t)$, $E_{\rm AGNR}(t)$ and $E_{\rm mol}$ are the total energies of the AGNR-molecule system, the graphene nanoribbon, and the chiral molecule at each considered time. Similarly, inspired by the recent concept of chemical bonding based on the isotropic molecular polarisability $\alpha$,\cite{Hait2023,Puleva2025}  the change in $\alpha$ was calculated according to: 
\begin{equation} \label{eq:Pol}
\Delta \alpha(t) = \alpha_{\rm sys}(t) - \alpha_{\rm AGNR}(t) - \alpha_{\rm mol}(t),
\end{equation}
\noindent with meanings similar to those indicated in Eq. \ref{eq:Eb}.
Finally, we considered the separation between the center of mass of the graphene nanoribbon and the chiral molecule, $\Delta d$: 
\begin{equation} \label{eq:dist}
\Delta d (t) = \left\| \frac{1}{N_{\text{mol}}} \sum_{j=1}^{N_{\text{mol}}} \mathbf{r}_{\text{mol},j}(t) - \frac{1}{N_{\text{AGNR}}} \sum_{k=1}^{N_{\text{AGNR}}} \mathbf{r}_{\text{AGNR},k}(t) \right\|.
\end{equation}
Here, \( N_{\text{mol}} \) and \( N_{\text{AGNR}} \) are the number of atoms in the chiral molecule and graphene nanoribbon, respectively. \( \mathbf{r}_{\text{mol},j} \) and \( \mathbf{r}_{\text{AGNR},k} \) are the position vectors of the $j$-th atom in the molecule and the $k$-th atom in the graphene nanoribbon. 

\section{Results and Discussion}

In this section, we will address the question formulated in the introduction, namely, under which conditions a chirality- "blind" substrate may be able to discriminate enantiomers on the basis of its electrical response. For this, we first analyze  differences in the electrical response between  enantiomer pairs for static geometries. Then, in a second step, quantum MD simulations are performed to account for the influence of structural fluctuations on the electrical response. Finally, potential applications of quantum mechanical-based features for chiral discrimination are discussed.

\subsection{Chiral discrimination: static case}

We first analyze the results of the electronic transport properties for optimized AGNR-molecule systems.
To illustrate the effect of chirality (L- and D-enantiomers) on the transport properties, we  consider the cases of aspartate (Asp) and valine (Val), see Figs. \ref{fig:2}(a,b).
The results for other chiral molecules can be found in Fig. S1 of the Supplementary Material (SM).
The comparison of the zero-bias transmission functions, $\tau(E, V_{DS}=0)$, for the L- and D-enantiomers of both molecules indicates almost negligible  deviations from the transmission function of the pristine graphene nanoribbon.
Consequently, no significant chiral sensitivity is observed in the linear conductance of the system, suggesting the necessity of nonequilibrium transport conditions to assess chiral effects in this device. 

To demonstrate this, Figs.{\ref{fig:2}}(c,d) display the difference in $\tau(E, V_{DS})$ of the corresponding enantiomers, $\Delta \tau = \tau_L-\tau_D$, for Asp and Val under nonequilibrium transport conditions (i.e., $V_{DS}\neq 0$).
We only present the results for the largest applied voltages $V_S = -V_D =$ 0.7 V (green solid line) and $V_S = V_D =$ -0.7 V (yellow solid line), where differences are most prominent.
This nonequilibrium transport calculation reveals relatively small, but already distinguishable differences between $\tau(E)$ of the L- and D-enantiomers. 
For Asp, the transmission at positive applied voltages is systematically larger for  L-Asp compared to  D-Asp ($\Delta \tau >0$), while the behavior becomes the opposite for a negative applied bias.  
In contrast, the situation is more involved for Val, which displays, at a given voltage, different trends in the sign of $\Delta \tau$ as a function of energy. 
As a result, the difference in currents, $\Delta I(V)=I_{\rm L}(V)-I_{\rm D}(V)$, will display different symmetries upon inversion of the voltage sign, see Fig.{\ref{fig:2}}(e), where this quantity is shown for all amino acids as a function of the applied voltage $V_S = -V_D \in [-0.7, 0.7] V$: $\Delta I(V)\approx -\Delta I(-V)$ for Val and Tyr, but $\Delta I(V)\approx\Delta I(-V)$ for the remaining amino acids. 
The current differences remain below 80 nA for the molecules, except for Asp, where it  reaches a few hundred nA at maximum bias (right scale).  
Phe exhibits the lowest values, while Tyr, with its hydroxyl termination, shows slightly higher values, suggesting a possible influence of the \ch{OH} group on chiral sensitivity. 
The distinct $\Delta I(V)$ responses indicate that molecular chemistry plays a key role in the sensing mechanism. 
In general terms, current differences in this order of magnitude are measurable, thus highlighting the importance of addressing a nonequilibrium transport regime.  
However, we remark that because of the small size of the chiral molecules, a single molecular configuration as discussed in this section may not be fully representative of the molecular conformational space and, thus, the obtained current-voltage characteristics may also not be representative either. Moreover, would the chiral discrimination power of the proposed setup be robust against the inclusion of structural fluctuations? To further explore these issues, we address in the next section the influence of the conformational dynamics at room temperature on the electrical response. 

\subsection{Chiral discrimination: the influence of the structural dynamics}

To allow the AGNR-molecule system to explore a larger conformational space and examine the possible implications for chiral discrimination, quantum Molecular Dynamics simulations at 300 K were performed using the DFTB3 method supplemented with an MBD treatment for modeling vdW interactions (see Sec. \ref{Sec:md} for more details). 
The analysis of the structural and transport properties was conducted using $160$ conformations from the last 8 ps of each MD simulation (out of a 10 ps long trajectory), during which we did not observe significant fluctuations in total energy or temperature (see Figs. S4-5 of the SI).
Figs. \ref{fig:3}(a,b) highlight the structural fluctuations (shadowed background) of both enantiomers of Asp and Val along the  MD trajectory, revealing significant conformational deviations from the optimized structure of the amino acids.
To quantify these structural changes, we have calculated the time evolution of the center-of-mass distance, $\Delta d$, between the amino acids and the graphene nanoribbon (see Eq. \ref{eq:dist}). 
Typical time series are shown in panels (c) and (d) of Fig. \ref{fig:3} for Asp and Val, respectively. 
In the former case, the fluctuations of $\Delta d$ are larger for D-Asp compared to  L-Asp, however, the average distances are circa 4.0 \AA{} for both enantiomers. 
In contrast, there is a clear difference in the fluctuations and average distances of  L-Val and D-Val, which can be attributed to different electrostatic and dispersion interactions governing the molecular dynamics.
A summary of the time series statistics for all amino acids is presented in Figs. \ref{fig:3}(e,f) in the form of boxplots for D- and L-enantiomers, respectively. 
The average distance of the D-enantiomer of all chiral molecules (red lines) from the nanoribbon remains nearly the same ($\approx$4.0 \AA{}), while for the L-enantiomer, Val and Gln exhibit average distances of 4.9 \AA{} and 4.5 \AA{}, respectively, which are larger than those of the L-enantiomers of the other amino acids.
Notice that the main difference in $\Delta d$ for some amino acids (e.g., Asp, Tyr) and their respective enantiomers lies in the spread of the distance values (see gray bars), which may have implications for the transport properties.

\begin{figure}[t!]
 \centering
 \includegraphics[width=\columnwidth]{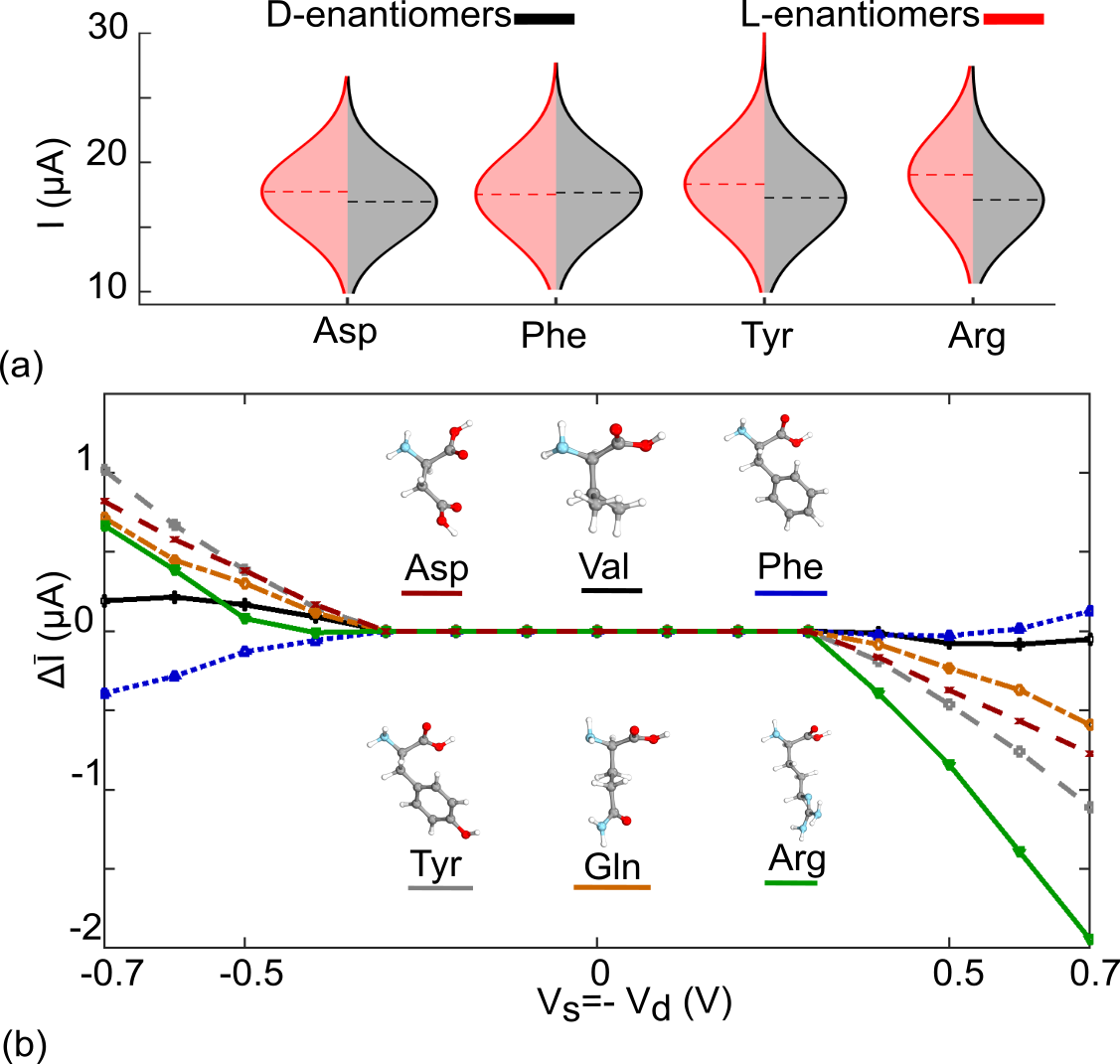}
 \caption{ (a) Distribution of electrical currents, $I$, obtained at an applied bias of $V_S = -V_D = 0.7$ V for Asp, Phe, Tyr, and Arg. 
(b) Variation of the difference in average electrical current, $\Delta \overline{I} = \overline{I}_L - \overline{I}_D$, as a function of applied voltage ($\Delta \overline{I}-V$ curves) for the non-equilibrium conformations of all analyzed amino acids.
To compute the properties presented in this figure, we have considered the conformations extracted from the last 8 ps of the MD trajectory.
}
 \label{fig:4}
\end{figure}

We then computed the charge current for the $160$ conformations of D- and L-enantiomers of a given chiral molecule.  
Fig. \ref{fig:4}(a) presents the distribution of current values, $I$, of Asp, Phe, Tyr, and Arg at an applied voltage $V_S = -V_D = 0.7 V$, with values for D-enantiomers in black and for L-enantiomers in red. 
Overall, the amino acids display a large variation of $I$ ranging from 10 to 30 $\mu$A, which is on the same order of magnitude as the values obtained for transport calculations on static structures (see Fig. S2 of the SM).
Moreover, we observe that L-enantiomers exhibit higher time average current $\overline{I}$ than their D counterparts, except for Phe, where both values are close to each other.
The corresponding time dependence of $I$ for all molecules can be found in Fig. S6 of the SM.
The effect of structural fluctuations on the chiral discrimination was also investigated for other applied voltages, and the results for  $\Delta \overline{I}=\overline{I}_L - \overline{I}_D$ are shown in Fig. \ref{fig:4}(b). 
All curves exhibit opposite $\Delta \overline{I}$ signs at the extreme bias values $V_S = \pm 0.7 V$. However, in the case of Phe, the sign is reversed, and $\overline{I}_L$ is smaller than $\overline{I}_D$ for $V_S = -0.7 V$. 
Concerning the $\Delta \overline{I}$ values, the chiral discrimination effect is largely enhanced by structural fluctuations, increasing from tens of nA obtained for the equilibrium conformations (see Fig. \ref{fig:1}(e)) to up to 1-2 $\mu$A when considering non-equilibrium conformations. 
This finding is a clear demonstration that dynamical effects are crucial for investigating chiral discrimination features of potential molecular chiral sensors. 
We have also found that Val and Phe show the smallest differences in averaged current between D- and L-enantiomers, which can be attributed to the lower chemical diversity in the achiral part of the molecules (consisting only of C-based chemical groups).
Although the $\Delta \overline{I}-V$ curves tend to display mirror symmetry with respect to the zero-bias point, Arg is the only amino acid that presents a considerable asymmetry between $\Delta \overline{I}$ values for positive and negative bias. This effect is related to  the complex electrostatic interactions between the N atoms in the guanidino group and the nanoribbon.

\begin{figure}[t!]
 \centering
 \includegraphics[width=\columnwidth]{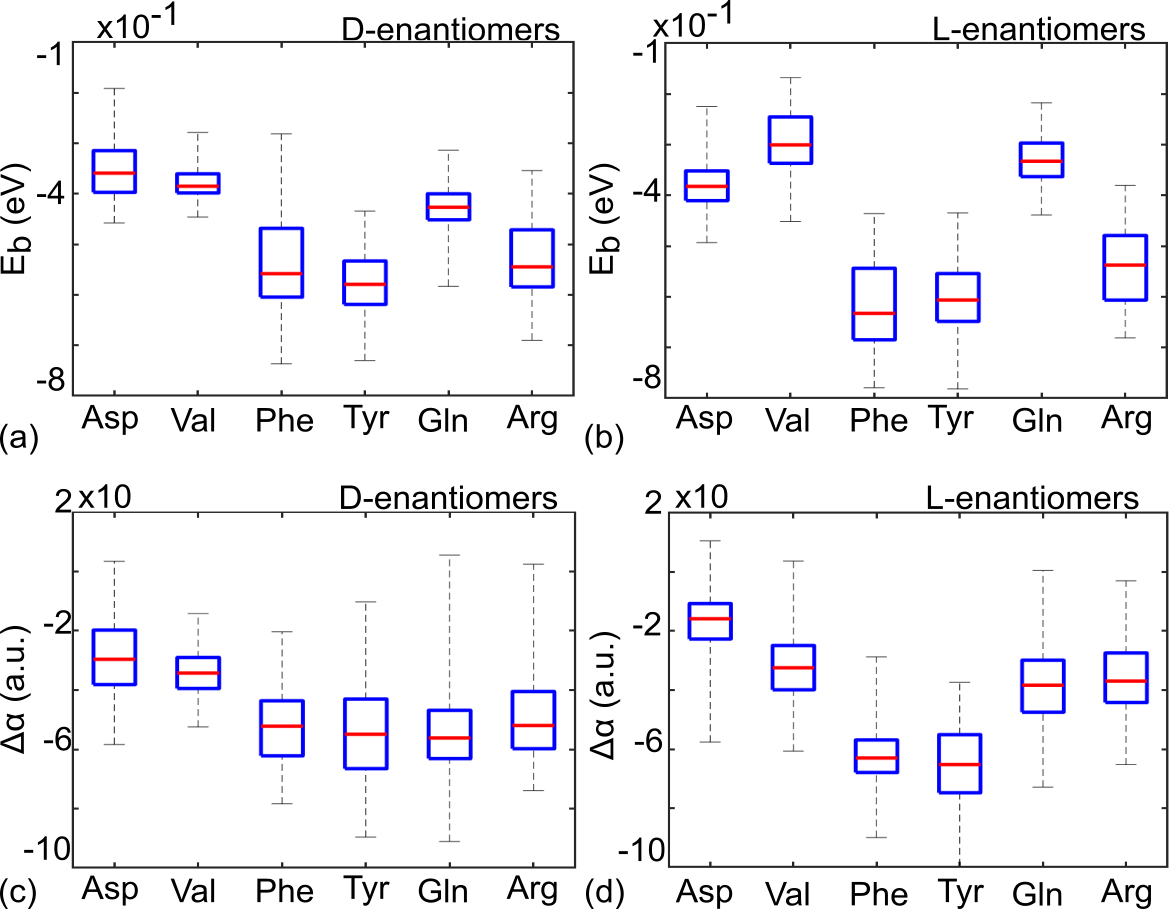}
 \caption{Boxplots of the binding energies, $E_b$, between the chiral molecule and AGNR15, obtained from the MD trajectory corresponding to (a) D-enantiomers and (b) L-enantiomers of all six amino acids.
Boxplots of the variation in molecular polarizability, $\Delta \alpha$ (see Eq. \ref{eq:Pol}), obtained during the MD trajectory corresponding to (c) D-enantiomers and (d) L-enantiomers of all six amino acids.
To compute the properties presented in this figure, we have considered the conformations extracted from the last 8 ps of the MD trajectory.
}
\label{fig:5}
\end{figure}

To further investigate the dynamics, we analyzed the parameters characterizing the interaction between the chiral molecules and the AGNR, starting with the binding energy, $E_b$ (see Figs. \ref{fig:5}(a,b)). 
The results show that amino acids containing a phenyl ring in their structures (i.e., Phe and Tyr) exhibit the largest $E_b$ values for both enantiomers due to the presence of $\pi-\pi$ interactions, followed by Arg, which contains a guanidino group.
Phe also presents the largest deviations from the average $E_b$ (red lines). However, the binding energies of the two enantiomers remain similar, suggesting comparable interactions with the nanoribbon.
As expected, Asp and Val present the weakest binding to the nanoribbon since they are the smallest and least complex chiral molecules considered in this study. 
The differences in $\Delta d$ values observed for D-Val and L-Val (see Figs. \ref{fig:3}(e,f)) are also reflected in the behavior of $E_b$, where D-Val has a larger average $E_b$ than L-Val. 

Additionally, to gain insights into the electrostatic interactions in GNR-molecule systems, we examined polarizability variations of the D- and L-enantiomers by computing $\Delta \alpha$, as defined in Eq. \ref{eq:Pol}. 
In Figs. \ref{fig:5}(c,d), one can see that $\Delta \alpha$ is sensitive to molecular composition and chirality.
For instance,  L-Phe and L-Tyr exhibit the largest variations in molecular polarizability ($\overline{\Delta \alpha}=$ -65 a.u.), while L-Asp and L-Val present small $\overline{\Delta \alpha}$ values such as -16 a.u. and -32 a.u., respectively.  
On the contrary, $\Delta \alpha$ values for D-enantiomers are closer, with average values ranging from -30 a.u. to -54 a.u., highlighting the potential of the parameter $\Delta \alpha$ to quantify chiral discrimination. 
In this context, there is an additional chirality-sensitive quantum mechanical quantity, the Rosenfeld tensor,\cite{10.1063/5.0227365} which can be incorporated into the characterization of the chiral molecule-substrate interaction.
These findings are in agreement with and complement the effects found in the analysis of $E_b$.

\begin{figure}[t!]
 \centering
 \includegraphics[width=\columnwidth]{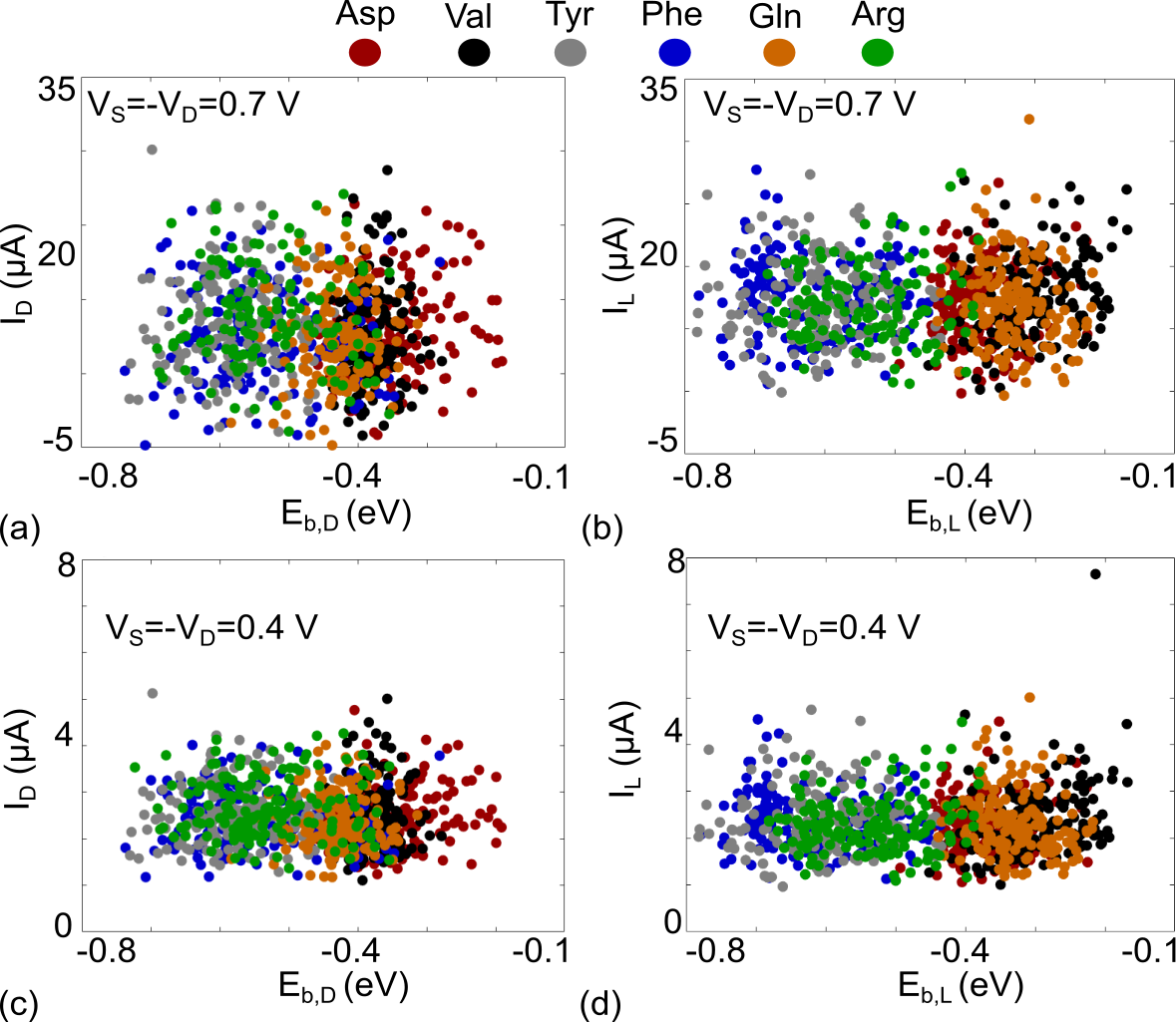}
 \caption{Correlation plots between the electrical current $I$ and the binding energy $E_b$ for (a,c) D-enanotiomers and (b,d) L-enantiomers of all six amino acids at two different applied biases $V_S=-V_D=$ 0.7 V and 0.4 V, respectively. 
The property values of each amino acid are represented by circles of different colors.
 %
To compute the properties presented in this figure, we have considered the conformations extracted from the last 8 ps of the MD trajectory.
 }
 \label{fig:6}
\end{figure}

We now examine the pairwise correlations between transport properties and binding features in the non-equilibrium conformations of both enantiomers of chiral molecules.
Fig. \ref{fig:6} shows the two-dimensional property spaces defined by the electrical current and the binding energy $\left( I,E_b \right)$ for the six amino acids at two different applied voltages: $V_S=-V_D=$ 0.4 V and 0.7 V. 
Overall, there is a lack of correlation between these properties for the whole set of conformations, with Pearson correlation factors $\rho<0.05$.  
By analyzing the $\left( I, E_b \right)$-space per chiral molecule, $\rho$ tends to increase for certain molecules and becomes dependent on the molecular composition and chirality (see Table S1 of the SM).
Indeed, at $V_S=0.7$, the enantiomers of Val--a molecule with small $\delta \overline{I}$ values-- present a large difference in $\rho$, i.e., $\rho=0.03$ for D-Val and $\rho=0.32$ for L-Val. 
A similar effect was observed for Asp and Gln enantiomers at both applied biases. 
Instead, the difference in $\rho$ values for the enantiomers of the other amino acids is small and less than 0.1.
It is worth noting that this lack of correlation is also evident when comparing electronic current values with polarizability variations, $\Delta \alpha$.
However, by analyzing the $\left( E_b, \Delta \alpha \right)$-space, we can identify a certain degree of correlation between these properties, supporting our discussion above (see Fig. S9 and Tables S2-3 of the SM).
A larger degree of correlation is obtained when analyzing $\left( E_b, \Delta d \right)$-space, with $\rho$ values ranging from 0.36 to 0.90 (see Fig. S9 and Tables S4-5 of the SM).
These property-property relationships provide a new mechanism for quantifying chiral discrimination effects in molecular sensors. 
The understanding of correlations in the property space spanned by chiral molecules can also aid in determining Pareto fronts for the multi-property optimization process of chiral sensors, as has been discussed for drug-like molecules.\cite{medrano23freedom,solvaqm} 

\section{Conclusions}
We have studied the conditions under which a chirality-"blind" sensing substrate can display enantioselectivity, using a small set of chiral amino acids interacting with a graphene nanoribbon as an illustrative example.
To tackle the problem, we applied the Landauer formalism, as implemented within the NEGF technique, in combination with the DFTB electronic structure method, including many-body dispersion corrections for van der Waals interactions. 
A first major result found from our simulations is that signatures of chiral discrimination appear only within a nonequilibrium transport regime; the linear response conductance (or transmission) of the system does not display any differences when exposing the nanoribbon to L- or D-enantiomers. 
A second major result is that the enantioselectivity of the investigated systems is enhanced when the structural dynamics of the active device region is included, leading to differences in currents for enantiomer pairs of up to $\approx$1-2 $\mu$A. 
Furthermore, we have proposed a new rationale for the dynamic effects in chiral discrimination through the analysis of binding features and property-property relationships. 
As a further investigation, a much larger pool of chiral molecules should be considered to explore not only the universality of the observed effects, but also to better identify unknown correlations between enantioselectivity and molecular properties.
Addressing this issue would demonstrate the "freedom of design"\cite{medrano23freedom,Szabolcs23} that exists in the property space of chiral molecules, enabling the discovery of potential molecular sensors for chiral discrimination---similar to the advancements made in olfactory molecular sensors.\cite{Chen2025} 
Hence, we expect our work to pave the way for advancing the development of machine learning-based computational frameworks aimed at understanding and designing efficient chiral sensors.

\begin{acknowledgments}
We acknowledge funding by the German Research Foundation (DFG) project "Data-Driven Characterization of (A)Chiral 2D Polymers" (CRC1415-C04). G.C. acknowledges funding by the ERA NET project "p-n Heterojunctions of emergent wide band gap oxides for self powered UVC sensing" (HEWOX, grant agreement ID: 01XX21006).
\end{acknowledgments}

\section*{Data Availability Statement}
The data that supports the findings of this study are available within the article and its supplementary material.

\appendix


\providecommand{\noopsort}[1]{}\providecommand{\singleletter}[1]{#1}%
%

\end{document}